\documentclass[pre,twocolumn,showpacs,floatfix,superscriptaddress]{revtex4}
\usepackage{graphicx,amsfonts,amssymb,amsmath, hyperref}
\usepackage[applemac]{inputenc}
\usepackage{multirow}

\newif\ifhyper
\hypertrue
\ifhyper
\hypersetup{
  citecolor = {green},
  colorlinks = {true}, 
  urlcolor = {blue} 
}
\fi

\newlength{\ldag}
\begin{document}

\title{Spontaneous symmetry breaking and the flat phase of crystalline membranes}

\author{O. Coquand} 
\email{coquand@lptmc.jussieu.fr}
\affiliation{Institut f\"ur Materialphysik im Weltraum, Deutsches Zentrum f\"ur Luft- und Raumfahrt (DLR), 51170 K\"oln, Germany}
\affiliation{Sorbonne Universit\'e, CNRS, Laboratoire de Physique Th\'eorique de la mati\`ere Condens\'ee, LPTMC, F-75005 Paris, France}


\begin{abstract}
	
	Crystalline membranes are one of the rare examples of bidimensional systems in which long-range order
	can stabilize an ordered phase in the thermodynamic limit.
	By a careful analysis of the Goldstone modes counting, we propose a symmetry-breaking mechanism
	associated with the generation of the flat phase, and we show how it highlights the crucial role played
	by the crystalline lattice in the establishment of long-range order in these objects.
	Comparison with other symmetry breaking mechanisms in membrane physics is also used to unveil the links 
	between symmetry breaking patterns and the physical properties of the flat phase.
	
\end{abstract}


\maketitle

	Crystals are generally defined as materials that possess an underlying periodic structure: at the
	microscopic scale, their atoms (or any other microscopic constituent) lie on a periodic lattice.
	However, this can only be true at $T=0$, otherwise thermal agitation causes the atoms to fluctuate
	around their equilibrium position -- these fluctuations are called phonons.
	For temperatures smaller than the melting temperature of the crystal, such fluctuations remain weak
	enough, so that each atom is still on average at its equilibrium position defined by the lattice.
	This is the definition of a crystalline phase at a non-zero temperature.
	
	This picture holds for most of the observed crystalline materials; however, it depends strongly on their
	dimension.
	In the 1930s, simple arguments were found, that show that crystalline order is indeed destroyed
	at any $T\neq0$ for one dimensional crystals in the thermodynamic limit \cite{Peierls34,Peierls35,Landau37}.
	Later, more rigorous arguments enabled the proof to be extended to two-dimensions \cite{Mermin68}.
	Thus, only in three dimensions can genuine crystals exist.
	
	In this context, the observation of graphene sheets -- a one-atom thick hexagonal lattice of carbon
	atoms -- first came as a surprise \cite{Novoselov04,Novoselov10}.
	Since then, many more bidimensional crystals have been observed and synthesized (see \cite{Roldan17}
	for example for a review).
	How does this agree with the previous findings about the impossibility of existence of crystalline phases
	in two dimensions ?
	It was later understood that height fluctuations (also called flexurons or flexural phonons),
	or more precisely their interaction with acoustic phonons,
	play a crucial role in maintaining the stability of such bidimensional crystal sheets:
	the long-range order is indeed not of a positional nature (as in the previous definition of crystals),
	but of an orientational one, i.e., it is not defined by the atoms being at precise positions on average,
	but by the vectors normal to the surface embodied by the crystal being correlated at large distances \cite{Aronovitz88,Aronovitz89}.
	Hence, despite their name, the ordered phase of crystalline membranes is not a crystalline phase, but a so-called
	flat phase (in reference to the membrane's typical configurations, although other geometries
	are allowed, e.g., as spherical shapes for vesicles).
	This can be seen quite clearly on diffraction patterns observed from graphene sheets \cite{Meyer07}.

	The presence of long range orientational order in crystalline membranes is at the origin of a number of unusual scaling behaviors,
	captured by an anomalous exponent $\eta$: the bending rigidity $\kappa$ increases with the size $L$ of the system as $\kappa\sim L^\eta$,
	whereas the elastic moduli such as Young's modulus $\mathcal{Y}$ get weakened as $\mathcal{Y}\sim L^{2-2\eta}$ (see \cite{Nelson04} and references therein).
	Additionaly, Hooke's law which determines the system's response to an external stress also becomes anomalous \cite{Aronovitz89,Guitter89,Gornyi17}.
	The anomalous exponent $\eta$ is moreover a universal quantity that does not depend on the nature of the material under consideration.
	
	Still, the theorem of Hohenberg-Mermin-Wagner \cite{Mermin66,Hohenberg67} forbids the existence of such an
	ordered phase in most two-dimensional systems.
	As for crystalline membranes, Nelson and Peliti have shown, in an attempt to
	examine the possibilities of a Kosterlitz-Thouless melting of the bidimensional crystals,
	that interactions between the Gaussian curvature of the membrane at different locations turn out
	to be long-range \cite{Nelson87}.
	In this paper we show that a similar line of reasoning can be applied at the level of the interaction involving the order parameter field
	which is also of long-range type, thereby providing an explanation of why the theorem of Hohenberg-Mermin-Wagner does not apply here.
	
	In the scenario of Hohenberg-Mermin-Wagner, long range order is destroyed by thermal fluctuations,
	which are dominated by the massless modes, if any, in the thermodynamic limit, because
	such modes do not get screened at large distances.
	The appearance of such modes generally requires particular symmetries to be at play in the system,
	which can prevent the existence of a mass term;
	it is then possible (at least in principle) to establish links between massless modes and the symmetries
	of the state in which the system lies.
	The Goldstone theorem \cite{Goldstone62} is one of the most useful tools that one can use to
	relate the symmetries of an ordered phase to the number of massless modes, but its original formulation
	is valid only for Lorentz-invariant systems.
	It has long been known that this theorem cannot be easily generalized to systems that do not
	possess this symmetry, the most well-known example being Heisenberg ferromagnets that have only one
	massless magnon instead of the two that would be expected from a naive application of Goldstone's theorem
	\cite{Nielsen76}.
	Despite thourough investigations (see \cite{Brauner10} for a review), a proper counting rule for the
	massless modes in full generality	only came out very recently
	\cite{Watanabe11,Watanabe12,Watanabe14,Hidaka13}.
	
	In this paper, we use these recent findings to shed new light on the puzzling question of long-range
	order in crystalline membranes.
	In particular, by applying the Goldstone counting rule \cite{Watanabe14}, we show that the previously proposed symmetry-breaking
	mechanisms associated with the flat phase \cite{Aronovitz89,Zanusso14,Guitter89} are not consistent with the well-known infrared spectrum of this
	phase because they forget the subtle (but nonetheless crucial) role played by the underlying crystalline lattice.
	Indeed, in membranes without such a lattice -- also called fluid membranes -- the
	long range orientational order is destroyed by thermal fluctuations \cite{DeGennes82}.
	The corrected mechanism is then compared with other symmetry breaking patterns from membrane physics
	to investigate further its relations to the theorem of Hohenberg-Mermin-Wagner.
	We identify two main important features of this mechanism (namely the occurrence of two
	different types of fluctuation modes and the presence of Goldstone modes with quadratic dispersion
	relation) that are, respectively, associated with
	the stabilization of the ordered phase in the thermodynamic limit and the generation of a non-trivial
	field anomalous dimension in the ordered phase, related to a strong degree of anharmonicity
	of the fluctuation modes, and which is a key quantity to understand the physics of the flat phase
	\cite{Nelson04}.
	
	The paper is organized as follows : We first present the Goldstone counting rule.
	Then, we apply it to find a consistent symmetry-breaking mechanism associated with the flat phase.
	In the third part, we show why long-range orientational order is preserved by performing a careful analysis of the
	hypotheses of Hohenberg-Mermin-Wagner's theorem.
	Finally, we apply the same procedure to the study of the overstretched phase, where the material undergoes a strong
	stretching effort, thereby giving another example
	that helps us track down the relations between the symmetry-breaking mechanism and the peculiar properties of
	the flat phase of crystalline membranes, and then we conclude.
	
\section{Goldstone modes counting without Lorentz invariance}

	Different low-energy states, that can be discriminated from each other, must have the same energy if
	they are related by a symmetry of the free energy.
	In particular, if a ground state breaks a symmetry of the free energy, all states that can be obtained
	by recursively applying this symmetry to the ground state are also possible ground states.
	This is how the comparative study of the symmetries of the states that the system can reach, and those
	of the free energy, gives insight into the degeneracies of the spectrum of the theory.
	In the following, we shall restrict ourselves to systems simple enough that more refined arguments
	about Higgs mechanisms, or other subtle ways to generate gaps \cite{Nicolis13,Nicolis13a} are not necessary.
	Then the low-energy spectrum can be directly read from the spontaneous symmetry-breaking pattern.
	
	Obviously, not all the symmetries need to be broken by the ground state, i.e., there can be residual symmetries.
	In the case when the broken symmetries are associated with continuous transformations, it is possible to
	relate any pair of ground states by a continous path of other ground states.
	That is, the spectrum includes massless excitations, which are the \textit{Goldstone modes}.
	Because of their massless nature, Goldstone modes are not secreened at large distances, and therefore they
	play a crucial role in the infrared physics of the system.
	In relativistic systems, Lorentz symmetry imposes on these modes a dispersion relation of the form
	$\omega=q c$ ; however this no longer holds for non-relativistic systems.
	For example, the ferromagnetic magnon has a low-energy dispersion relation of the form $\omega\sim q^2$
	\cite{Nielsen76}.
	Hence for systems without Lorentz invariance, the counting rule should give access not only to the number
	of Goldstone modes, but also to their type of dispersion relation.
	
\begin{figure}
		\begin{center}
			\includegraphics[scale=1.35]{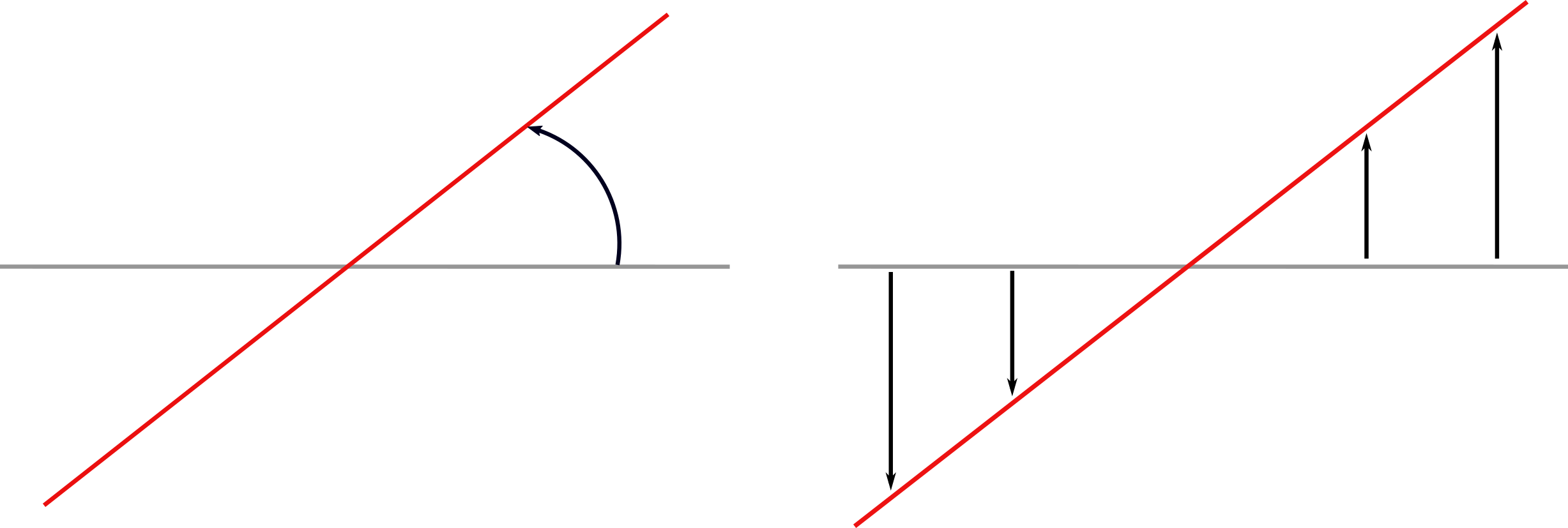}
		\end{center}
		\caption{Left: effect of a rotation on a line. Right: the action, on the same line, of a combination
		of local translations induces the exact same transformation of the line. It is one of the simplest
		examples in which mathematically independent transformations can induce similar excitations
		of a given state.
		This example is discussed in \cite{Low02}.}
		\label{figLM}
	\end{figure}

	The counting of Goldstone modes can be related to the algebra of the symmetry groups of the
	free energy and the ground states by Goldstone's theorem \cite{Goldstone62} : let $G$ be the
	symmetry group of the free energy, and let $H$ be the group of symmetries of the ground states, namely
	the group of residual symmetries, with $H\subset G$ in light of the above.
	Goldstone's theorem (in its original version) states that the number of Goldstone modes is
	simply given by dim$(G/H)$.
	However, we must note the following:
	\begin{itemize}
		\item this rule cannot be trivially extended to non-relativistic systems.
		Indeed, in ferromagnets for example, $G=$ O$(3)$ is broken into $H=$ O$(2)$ (the ground state is still
		invariant with respect to rotations of axis in the direction of the spontaneous magnetization).
		The ferromagnetic state thus breaks two generators of $G$, but there is only one magnon
		\cite{Nielsen76}.
	
		\item As Goldstone's theorem is intended only for relativistic systems, it does not give any
		information about the dispersion relations, which are crucial characteristics of the massless
		modes in non-relativistic theories.
		It was first pointed out by Nielsen and Chadha that the infrared behavior of the Goldstone bosons
		has a direct influence on the number of generated massless modes :
		to continue with the example of magnets, ferromagnets have one magnon with a quadratic dispersion
		relation $\omega\sim q^2$ , whereas antiferromagnets have two magnons with a linear dispersion
		relation $\omega\sim q$ , while in both cases two rotation generators are broken by the ground
		state \cite{Nielsen76}.
		
		\item the knowledge of the algebra of broken generators is not sufficient in general.
		Indeed, two different broken generators can generate the same excitation on a given
		ground state (see Fig. \ref{figLM} for a simple example), and therefore they are not associated
		with different Goldstone modes \cite{Low02}.
		This is all the more important for systems that possess spacetime symmetries which frequently
		generate non-independent transformations of the ground states \cite{Kharuk18}.
		
	\end{itemize}
	
	It can be understood in light of the above that the central quantity involved in the counting of
	Goldstone bosons should include more information than the mere algebra of the broken generators.
	Let us define the matrix $\rho$ of the \textit{commutators} of \textit{independent} broken generators
	$\left\{Q_i\right\}$ \textit{evaluated} in the ground state $\left|0\right>$ \cite{Watanabe12} :
	\begin{equation}
		\rho_{ab}=\left<[Q_a,Q_b]\right>\ .
	\end{equation}
	It is also important to discriminate the generated massless modes according to their dispersion relation:
	we will call type-A Goldstone bosons those that have a linear dispersion relation, and type-B those
	whose dispersion relation is quadratic (the rigorous definition is a little bit more subtle,
	but the difference is irrelevant to our purpose \cite{Watanabe14}).
	The numbers $n_A$ and $n_B$ of each type of Goldstone modes is then given by the following formulas
	\cite{Watanabe12}:
	\begin{equation}
	\label{eqWata}
		\left\{\begin{split}
			&n_A = \text{dim}(G/H)-\text{rank}(\rho)\\
			&n_B = \frac{\text{rank}(\rho)}{2}\ .
		\end{split}\right.
	\end{equation}
	
	Note that in the relativistic case, the existence of a non zero expectation value of the commutator
	in the ground state would break Lorentz invariance \cite{Watanabe14}, thus Lorentz symmetry requires
	$\rho=0$, and we recover the original Goldstone's theorem with dim$(G/H)$ type-A massless modes.
	
	As a less obvious example, consider an ideal crystalline solid in $D$ dimensions.
	The ground state of the system is given by the periodic lattice that breaks both the translation
	and rotation symmetries of the free-energy (given in that case by the theory of elasticity \cite{Landau90}).
	The residual symmetry group $H$ is the discrete subgroup of symmetry of the crystalline lattice
	denoted $\mathcal{C}$.
	The spontaneous symmetry breaking mechanism characterizing the crystalline ground state can be
	written as follows:
	\begin{equation}
	\label{eqM1}
		\text{\underline{\textbf{Mechanism 1:}}} \quad \text{ISO}(D)\rightarrow \mathcal{C}\ ,
	\end{equation}
	where ISO$(D)$ is the group of isometries.
	
	There are $D$ broken translation generators, as well as $D(D-1)/2$ broken rotation generators.
	However, the action of translations and rotations on the ground state do not generate independent
	excitations \cite{Watanabe13,Beekman17a}, and the group of \textit{independent} broken generators reduces to the
	broken translations.
	Because translations commute with each other, $\rho=0$ and the counting rule Eq.(\ref{eqWata}) gives
	$D$ type-A Goldstone bosons, corresponding to the well-known $D$ acoustic phonons of crystals, and not
	$D+D(D-1)/2=D(D+1)/2$ Goldstone bosons as a naive application of Goldstone's theorem would have
	given.
	
	This simple example sheds light onto two properties of the counting rule Eq.(\ref{eqWata}).
	First, for non-Lorentz-invariant systems, the total number of Goldstone modes does not need to be equal to the total
	number of broken generators, even if all of them have a linear dispersion relation (the equality occurs only
	if we suppose further that the actions of each of the generators on the ground state are independent).
	Second a non-trivial algebra of independent broken generators is required for the existence of type-B
	Goldstone bosons.	 

\section{Spontaneous symmetry breaking in the flat phase}
\label{secII}

	In this section, we study the application of the Goldstone counting rule Eq. (\ref{eqWata}) to the flat phase.
	Note that we are concerned here with the symmetries of the zero-temperature ground state of the crystalline membrane,
	which is still a periodic crystal.
	The influence of thermal fluctuations on the stability of such a state is discussed in the next section.

	In addition to usual crystals, a fundamental property of crystalline membranes is their ability
	to fluctuate in a bigger embedding space, whose dimension will be called $d$.
	This paves the way to more complex spacetime symmetry breaking patterns, which, as we can already
	anticipate will be of paramount importance to explain the quadratic dispersion relation of flexurons
	(see below).
	The massless fluctuation spectrum of the flat phase of crystalline membranes is well-known.
	It includes the following\cite{Nelson04}:
	\begin{itemize}
		\item $D$ acoustic phonons, which are type-A Goldstone bosons, as in any crystal.
		
		\item $d-D$ flexurons which are type-B Goldstone bosons, and therefore much stronger than
		the phonons in the infrared limit.
		
	\end{itemize}

	There is no clear consensus in the literature on the spontaneous symmetry-breaking pattern
	associated with the flat phase to the best of our knowledge \cite{Aronovitz89,Zanusso14,Guitter89}.
	Note that most of these patterns were proposed at a time when the Goldstone counting rule was not known.
	
\begin{figure}
		\begin{center}
			\includegraphics[scale=1]{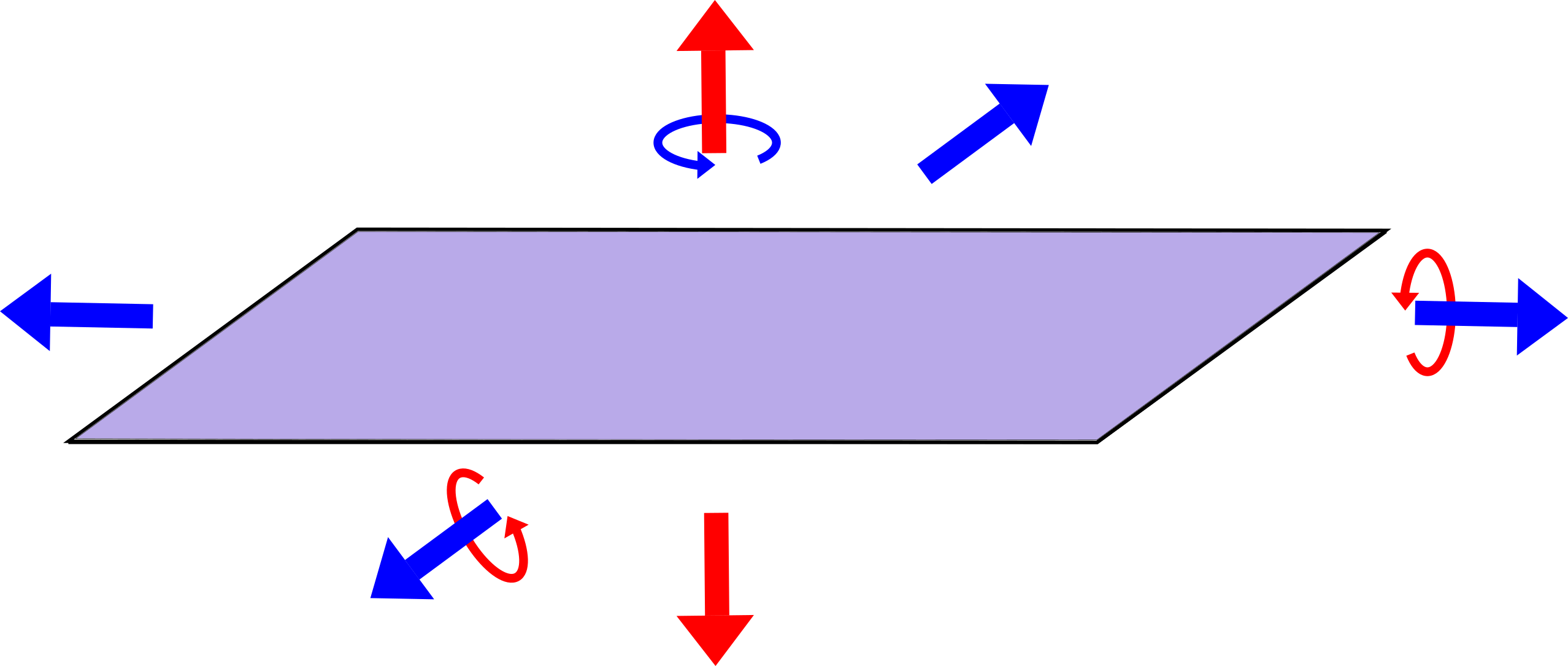}
		\end{center}
		\caption{Symmetries of the infinite plane in three dimensions : the plane is invariant under
		two translations and one rotation (in blue in the figure), but it breaks the other two rotations and the
		remaining translation (in red).
		The SO$(d-D)$ group reduces to the discrete $\mathbb{Z}_2$ group corresponding to reversing
		the up and down sides of the plane.}
		\label{figSymPlan}
	\end{figure}

	In the following we will only discuss the more widely used mechanism, which is also the one that
	best respects the symmetries of the flat phase (see Fig. \ref{figSymPlan}) and argue why we think 
	it is not correct.
	Consider the following symmetry breaking pattern \cite{Guitter89}:
	\begin{equation}
	\label{eqM2}
		\text{\underline{\textbf{Mechanism 2:}}} \quad \text{ISO}(d)\rightarrow\text{ISO}(D)\times
		\text{SO}(d-D)\ .
	\end{equation}
	
	The free-energy is invariant under all isometries of the embedding space, forming the group ISO$(d)$,
	and the flat phase configuration is an infinite plane, which is still invariant under the plane isometries
	ISO$(D)$ as well as the rotations of SO$(d-D)$ which act only on directions of the embedding space 
	that are all orthogonal to the flat phase plane (see Fig. \ref{figSymPlan} for a picture for the physical
	dimensions $D=2$, $d=3$).
	
	The set of broken symmetry generators thus contains the following:
	\begin{itemize}
		\item the $d-D$ translations in the directions orthogonal to the flat phase plane, denoted
		$\left\{P_\alpha\right\}_{\alpha\in[\![D+1;d]\!]}$.
		In the following, greek letters denote indices in $[\![1;d]\!]$, whereas latin ones only run
		into $[\![1;D]\!]$.
		
		\item the $D\times(d-D)$ \textit{mixed} rotations that bring the vector giving the $i^{th}$ direction
		inside the flat phase plane to the direction given by the $\alpha^{th}$ vector of the embedding
		space outside the membrane's plane $\left\{J_{i\alpha}\right\}_{i\in[\![1;D]\!]
		,\alpha\in[\![D+1;d]\!]}$.
		
	\end{itemize}
	
	The commutation relations between these generators are given by the $\text{iso}(d)$ algebra of
	the isometry group \cite{Schwarz71}:
	\begin{equation}
	\label{eqisod}
		\begin{split}
			[J_{\mu\nu},J_{\alpha\beta}]&=\delta_{\nu\alpha}J_{\mu\beta}-\delta_{\mu\alpha}J_{\nu\beta}
			-\delta_{\nu\beta}J_{\mu\alpha}+\delta_{\mu\beta}J_{\nu\alpha}\\[0.2cm]
			[J_{\mu\nu},P_{\theta}]\ \,&=\delta_{\nu\theta}P_{\mu}-\delta_{\mu\theta}P_{\nu}\\[0.2cm]
			[P_\mu,P_\nu]\ \ \,&=0\ .
		\end{split}
	\end{equation}
	The next question one should answer is which of these generators act independently on the ground state
	$\left|0\right>$.
	Consider the action of a mixed rotation on the ground state :
	\begin{equation}
		J_{i\alpha}\left|0\right>=\big(x_iP_\alpha-x_\alpha P_i\big)\left|0\right>
		=x_i P_\alpha\left|0\right>\ .
	\end{equation}
	The second equality follows from the fact that the translations inside the flat phase plane are not broken.
	Hence the action of a mixed rotation can always be canceled by a combination of broken translations.
	The set of independent broken generators thus reduces to $\left\{P_\alpha\right\}_{\alpha\in[\![D+1;d]\!]}$,
	and the counting rule Eq.(\ref{eqWata}) yields:
	\begin{equation}
	\label{eqG2}
		\left\{\begin{split}
			&n_A=d-D\\
			&n_B=0\ ,
		\end{split}\right.
	\end{equation}
	which is not consistent with the expected spectrum for the flat phase of crystalline membranes.
	
	This computation is quite enlightening though, since it underlines the necessity of having a non-trivial
	algebra of broken generators to describe type-B Goldstone bosons such as the flexurons in
	crystalline membranes.
	Looking more thoroughly at Eq. (\ref{eqG2}), we notice that the number of generated Goldstone bosons
	is equal to the number of directions of the embedding space orthogonal to the flat phase plane, which
	indicates that they must be flexurons (but not with the expected dispersion relation): the
	mechanism in Eq. (\ref{eqM2}) misses the phonons.
	Thus, this mechanism seems more appropriate to describe the flat phase of fluid membranes
	\footnote{Such an ordered phase only exists at $T=0$ however, because of Hohenberg-Mermin-Wagner's
	theorem.}
	in which only flexurons are at play.
	As a matter of fact, the constituents of an incompressible fluid membrane are free to diffuse on its surface,
	thereby forbidding the definition of any particular reference state by the position of
	the molecules, and therefore any kind of positional order.
	Such materials thus do not have acoustic phonons.

	This assertion can indeed be checked for consistency with usual models of fluid membranes:
	from the Canham-Helfrich free-energy \cite{Canham70,Helfrich73}, the flexuron's propagator $G(q)$
	has the following asymptotic behavior in the infrared limit:
	\begin{equation}
	\label{eqh}
		G(q)=\left<h(q)h(-q)\right>\underset{q\rightarrow0}{\sim}\frac{1}{\sigma q^2}\ ,
	\end{equation}
	where $\sigma$ is the tension of the membrane.
	Such behavior is typical of type-A Goldstone bosons.
	One could argue that, according to Eq.(\ref{eqh}), another type of behavior could be expected in tensionless
	fluid membranes, but this does not hold since even if not present at the microscopic scale, $\sigma$
	is generated by the renormalization group flow when going towards the infrared regime \cite{Nelson04}.
	Our Goldstone modes analysis leads to the following complementary argument: the tension term is not
	protected by the symmetries of the system, and therefore cannot be consistently enforced to be zero.
	
	As we have seen in Eq.(\ref{eqM1}), the origin of acoustic phonons in crystals is the breaking
	of the isometry group by the discrete group of the crystalline lattice $\mathcal{C}$.
	This must also hold for crystalline membranes in which, although not well preserved by the thermal
	fluctuations, a crystalline lattice is still present in the flat phase.
	In light of these arguments, it seems reasonable to take a closer look at the following
	symmetry-breaking pattern:
	\begin{equation}
	\label{eqM3}
		\text{\underline{\textbf{Mechanism 3:}}} \quad \text{ISO}(d)\rightarrow\mathcal{C}\times
		\text{SO}(d-D)\ .
	\end{equation}
	
	Note that the track of the discrete symmetry group $\mathcal{C}$ is still present even in the continuum
	theory of crystalline membranes:
	in \cite{David88} for example, the transition to the flat phase is presented as the generation of a non-zero
	expectation value for the metric of the flat phase $g_{ij}^0$ which then characterizes the ground state.
	Because the membrane is assumed to be homogeneous and isotropic, its metric in the ground state
	has to be proportional to $\delta_{ij}$.
	The proportionality coefficient $\zeta^2$, called the extension parameter, characterizes the unit
	of length inside the membrane's plane.
	The presence of this fixed reference metric is one of the key differences between crystalline membranes
	and fluid ones.
	
	The broken symmetry generators in the mechanism given by Eq.(\ref{eqM3}) are thus as follows:
	\begin{itemize}
		\item the $d-D$ external translations $\left\{P_\alpha\right\}_{\alpha\in[\![D+1;d]\!]}$.
		
		\item the $D$ internal translations $\left\{P_i\right\}_{i\in[\![1;D]\!]}$.
		
		\item the $D\times(d-D)$ mixed rotations, mixing the internal and external spaces
		$\left\{J_{i\alpha}\right\}_{i\in[\![1;D]\!],\alpha\in[\![D+1;d]\!]}$.
		
		\item the $\frac{D(D+1)}{2}$ internal rotations $\left\{J_{ij}\right\}_{(i,j)\in[\![1;D]\!]^2}$.
		
	\end{itemize}
	
\begin{figure}
		\begin{center}
			\includegraphics[scale=1.35]{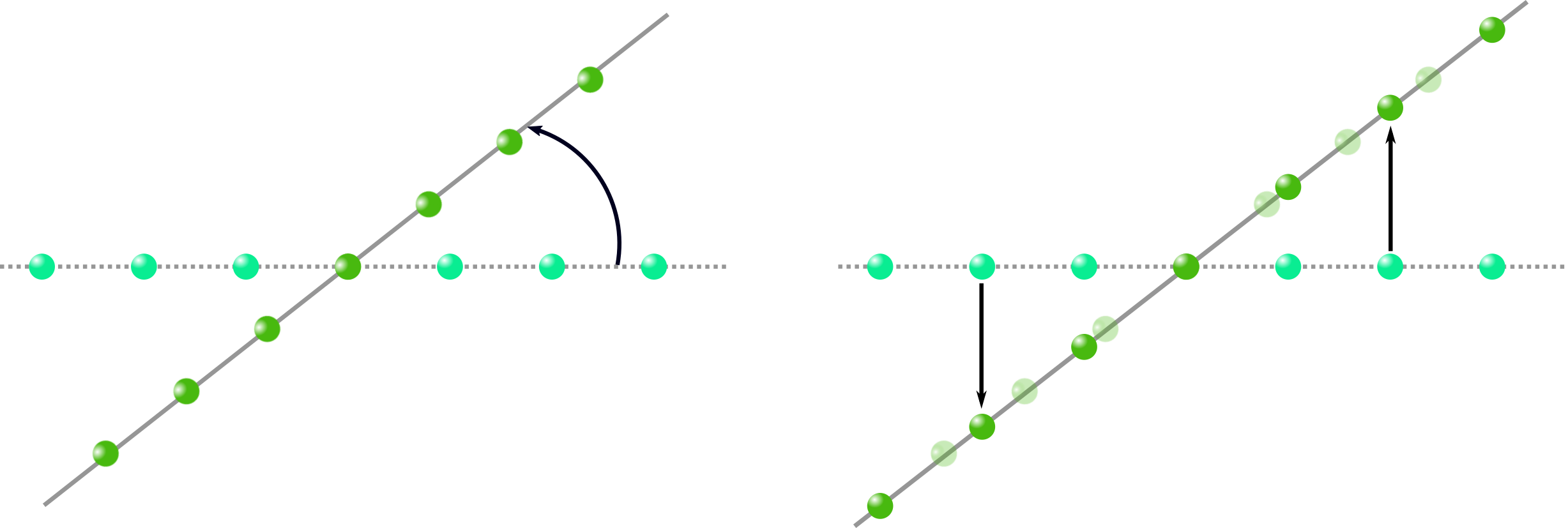}
		\end{center}
		\caption{Left: action of a mixed rotation on a one-dimensional lattice.
		Right: a linear combination of translations can align the lattice on the same line,
		but the state of the system is different because of the dilation that the translations induce
		on the lattice due to Pythagoras' theorem.}
		\label{figRotMix}
	\end{figure}

	The action of internal translations and rotations are not independent, as in usual crystals.
	This time however, the action of mixed rotations on the ground state can no longer be canceled by a
	carefully chosen combination of external translations
	(see Fig. \ref{figRotMix} for a picture in the one dimensional case).
	Indeed, whereas rotations always preserve the ground state metric, a combination of translations
	bringing the system into a similar plane would induce a dilation of the system (as an obvious
	application of Pythagoras' theorem shows) and therefore a flat phase state with a different extension
	parameter $\zeta$.
	This is a direct consequence of the presence of a microscopic lattice, or equivalently of the presence
	of phonons in the system.
	
	The internal space isometries can be used to relate the mixed rotations with the same external index
	$\alpha$, but different internal indices $i$ .
	All in all, the total number of independently acting broken symmetry generators is as follows:
	$D$ internal translations $+$ $(d-D)$ external translations $+$ $(d-D)$ mixed rotations,
	so that finally:
	\begin{equation}
		\text{dim}(G/H)=2d-D\ .
	\end{equation}
	
	The commutations relations between these generators evaluated in a ground state are given as 
	usual by the algebra of isometries $\text{iso}(d)$ :
	\begin{equation}
	\label{eqisod2}
		\begin{split}
			&\left<[J_{\alpha i},J_{\beta i}]\right> \propto\left<J_{\alpha\beta}\right>=0\\
			&\left<[P_{\alpha},P_{\beta}]\right>=0\\
			&\left<[J_{\alpha i},P_{\gamma}]\right>\underset{\gamma\neq i,\gamma \neq\alpha}{=}0\\
			&\left<[J_{\alpha i},P_{\alpha}]\right>=\left<P_i\right>\neq0\\
			&\left<[J_{\alpha i},P_{i}]\right>=\left<P_\alpha\right>\neq0\ .
		\end{split}
	\end{equation}
	It is then possible to write $\rho$ is a basis where the determination of its rank is simple,
	even without knowing the precise value of the non-zero matrix coefficients, denoted "$*$" below.
	We choose to write in a basis in which  the $d$ translations are displayed first, and then
	the $d-D$ rotations :
	\begin{equation}
		\rho=\left[
		\begin{array}{ccccc|ccccc}
			0      & 0   & 0   & 0   & ... & *   & *   & 0   & 0   & ... \\
			0      & 0   & 0   & 0   & ... & *   & 0   & *   & 0   & ... \\
			0      & 0   & 0   & 0   & ... & *   & 0   & 0   & *   & ... \\
			0      & 0   & 0   & 0   & ... & *   & 0   & 0   & 0   & ... \\
			...    & ... & ... & ... & ... & ... & ... & ... & ... & ... \\
			\hline
			*      & *   & *   & *   & ... & 0   & 0   & 0   & 0   & ... \\
			*      & 0   & 0   & 0   & ... & 0   & 0   & 0   & 0   & ... \\
			0      & *   & 0   & 0   & ... & 0   & 0   & 0   & 0   & ... \\
			0      & 0   & *   & 0   & ... & 0   & 0   & 0   & 0   & ... \\
			...    & ... & ... & ... & ... & ... & ... & ... & ... & ...
		\end{array}\right]\ ,
	\end{equation}
	which leads to:
	\begin{equation}
		\text{rank}(\rho)=2(d-D)\ ,
	\end{equation}
	and finally, thanks to the counting rule Eq. (\ref{eqWata}),
	\begin{equation}
	\label{eqG3}
		\left\{\begin{split}
			&n_A=D\\
			&n_B=d-D\ .
		\end{split}\right.
	\end{equation}
	Finally, we get $D$ type-A Goldstone modes with linear dispersion relation, corresponding to the
	acoustic phonons, and $d-D$ type-B Goldstone bosons with quadratic dispersion relation
	corresponding to the flexurons; which is the exact spectrum of crystalline membranes that we
	recalled at the beginning of the section.
	
	The crucial difference between the second and third proposed mechanisms -- Eq. (\ref{eqM2}) and
	Eq. (\ref{eqM3}) respectively -- is the presence of the crystalline lattice discrete group in the
	latter, that allows to keep some independent rotation generators required to get a non trivial
	algebra Eq. (\ref{eqisod2}) and thus a non zero rank for $\rho$, which allows for the existence
	of the type-B flexurons.
	The consequences are far-reaching as among the proposed mechanisms, only the third one is associated
	with a stable ordered phase in two dimensions: phonons Eq. (\ref{eqM1}) or flexurons Eq. (\ref{eqM2})
	alone cannot stabilize a long range order of positional or orientational nature.
	Already at this stage, we can notice two main differences between mechanisms 1 and 2 on the
	one hand and mechanism 3 Eq. (\ref{eqM3}) on the other hand (although they are all three built
	from the same groups), namely that mechanism 3 has two different types of fluctuation modes,
	and it has type-B Goldstone modes.
	The rest of this paper is dedicated to the analysis of the consequences of these two differences.

\section{Hohenberg-Mermin-Wagner's theorem}

	The spontaneous symmetry breaking pattern combined with the counting rule Eq.(\ref{eqWata}) gives access
	to the number of massless modes as well as their dispersion relation in the large distance limit in the
	ordered phase predicted by mean-field theory, but it is not sufficient to know if such an ordered
	phase is robust to thermal fluctuations.
	For that last purpose, the most useful tool is the Hohenberg-Mermin-Wagner's theorem
	\cite{Hohenberg67,Mermin66}.
	In their original paper, Mermin and Wagner stress an important hypothesis for their theorem to apply:
	the interaction needs to have a short enough range.
	Namely, if $J$ is the coupling constant describing the strength of the interaction between the
	order parameter fields, $J(x)x^2$ must be an integrable function in $D$ dimensions \cite{Mermin66}.
	Let us test this hypothesis in the case of crystalline membranes.

	First, we need the action describing the small fluctuations around the flat phase.
	As stated before, crystalline membranes can be described as an elastic medium fluctuating in an embedding space, and
	their action thus contains both a curvature term, proportional to the bending energy $\kappa$ of the
	membrane, and an elastic term quadratic in the strain tensor $\varepsilon_{ij}$:
	\begin{equation}
		S=\int_x\left[\frac{\kappa}{2}\big(\partial^2\vec{r}\big)^2+\frac{c_{ijab}}{2}\varepsilon_{ij}
		\varepsilon_{ab}\right]\ ,
	\end{equation}
	where $\int_x=\int d^Dx$ is an integral over the internal space of the membrane, $\vec{r}$ is the
	position vector describing the membrane, and
	$c_{ijab}$ is the elastic tensor, which can be expressed for example in terms of the Lam\'e coefficients
	as $\lambda \delta_{ij}\delta_{ab}+\mu(\delta_{ia}\delta_{jb}+\delta_{ib}\delta_{ja})$.
	The strain tensor $\varepsilon_{ij}$ can be expressed as half the difference between the metric in the current state of the membrane 
	$g_{ij}=\partial_i\vec{r}.\partial_j\vec{r}$ and that of the flat phase reference state
	$g_{ij}^0=\zeta^2\delta_{ij}$.
	To build a theory of small fluctuations around the flat phase, it is necessary to expand
	$\vec{r}$ around the equilibrium configuration with extension parameter $\zeta$ :
	\begin{equation}
		\vec{r}=\zeta x_i\vec{e}_i+\vec{u}+\vec{h}\ ,
	\end{equation}
	where a basis of the flat phase plane $\{\vec{e}_i\}$ has been introduced.
	This expansion causes the phonons $\vec{u}$ and the flexurons $\vec{h}$ to appear explicitely.
	Using the fact that the phonons and flexurons vibrate in orthogonal spaces, the action
	writes in terms of the most relevant terms reads \cite{Aronovitz88} :
	\begin{equation}
	\label{eqFl}
		\begin{split}
			S=\int_x\bigg[&\frac{\kappa}{2}\big(\partial^2\vec{h}\big)^2
			+\frac{c_{abcd}}{2}u_{ab}u_{cd}+\frac{c_{abcd}}{2}u_{ab}(\partial_c\vec{h}.\partial_d\vec{h})\\
			&+\frac{c_{abcd}}{8}(\partial_a\vec{h}.\partial_b\vec{h})(\partial_c\vec{h}.\partial_d\vec{h})\bigg]
			\ ,
		\end{split}
	\end{equation}
	where the symmetric tensor $u_{ab}=(\partial_au_b+\partial_bu_a)/2$ has been introduced.
	
	At first glance, it seems that the action Eq.(\ref{eqFl}) contains only local interactions between phonons
	and flexurons, what could lead one to conclude that the long-range orientational order is broken by
	thermal fluctuations.
	But in the original argument of Mermin and Wagner, there is only one fluctuation mode.
	
	We have seen that flexurons are the dominant modes in the infrared limit.
	Moreover, the action Eq.(\ref{eqFl}) is quadratic in the phonons, it is thus possible to perform an
	exact integration over the phonons to define an effective action with the flexurons as only fields
	\cite{LeDoussal92}.
	For sake of simplicity, we give it here in the Fourier space, with implicit momentum conservation,
	and the shorthand notation $\vec{h}(k_i)=\vec{h}_i$ :
	\begin{equation}
	\label{eqFeff}
		\begin{split}
			S_{\rm eff}&=\int_{k_1,k_2}\frac{\kappa}{2}\,k_1^4\big(\vec{h}_1.\vec{h}_2\big)\\
			+&\int_{k_1,k_2,k_3,k_4}\bigg[\frac{\mathcal{R}_{abcd}(q)}{4}k_1^ak_2^bk_3^ck_4^d\big(\vec{h}_1.\vec{h}_2\big)
			\big(\vec{h}_3.\vec{h}_4\big)\bigg]\ ,
		\end{split}
	\end{equation}
	with $\int_k=\int\frac{d^Dk}{(2\pi)^D}$	according to our convention for Fourier transforms, and $q=k_1+k_2=-k_3-k_4$.
	
	The price for working with one type of field only is that now the interaction vertex $\mathcal{R}$
	is non local.
	It depends only on two coupling constants, exactly like the elasticity tensor $c$, which can be
	made explicit by decomposing it onto the following set of orthogonal projectors:
	\begin{equation}
	\label{eqDefNM}
		\begin{split}
			& N_{abcd}(q) = \frac{1}{D-1}P^T_{ab}(q)P^T_{cd}(q)\\
			& M_{abcd}(q) = \frac{1}{2}\big(P^T_{ac}(q)P^T_{bd}(q)+P^T_{ad}(q)P^T_{bc}(q)\big)-N_{abcd}(q)\ ,
		\end{split}
	\end{equation}
	where $P^T_{ij}(q)=\delta_{ij}-q_iq_j/q^2$ is the projector in the direction orthogonal to $q$.
	The effective interaction vertex is then \cite{LeDoussal92} :
	\begin{equation}
	\label{eqReff}
		\mathcal{R}_{abcd}(q)=\frac{\mu(D \lambda+2\mu)}{\lambda+2\mu}N_{abcd}(q)+\mu M_{abcd}(q)\ .
	\end{equation}
	Note that in the particular case $D=2$, corresponding to physical membranes, the two projectors
	Eq.(\ref{eqDefNM}) are equal, and $\mathcal{R}$ depends on only one elastic constant, which turns
	out to be Young's modulus $\mathcal{Y}$ \cite{Nelson87}.

	The first proof of the presence of long-range content of the interaction in Eq.(\ref{eqFeff}) has been given by Nelson and Peliti \cite{Nelson87},
	they showed that in two dimensions, the interaction term in Eq.(\ref{eqFeff}) $S_{int}$ can be rewritten as an interaction between the
	local Gaussian curvature $s(x)=\text{det}(\partial_i\partial_j h)$ :
	\begin{equation}
		S_{int}=\frac{\mathcal{Y}}{16\pi}\int_{x,y}\mathcal{G}(x-y)s(x)s(y)\ ,
	\end{equation}
	where the (non-local) interaction vertex between the Gaussian curvature $\mathcal{G}$ behaves as $\mathcal{G}(x)\simeq x^2 \ln(x/a)$ at large distance
	($a$ being the lattice spacing), which is clearly a long-range type of interaction.

	To make the connection with the original work of Mermin and Wagner \cite{Mermin66}, we must first find the equivalent to the $J(x)$ interaction term.
	An order parameter associated to the flat phase is given by the extension parameter $\zeta$ which is always non zero in the flat phase, and equal to zero in a completely
	disordered crumpled configuration.
	It can also be expressed as a function of the corraltion between the tangent vectors to the surface generated by the membrane \cite{Guitter89}:
	\begin{equation}
		\zeta^2=\frac{1}{D}\left<\partial_i\vec{r}\right>.\left<\partial_i\vec{r}\right>\ .
	\end{equation}
	Finally, taking into account the fact that the action in Eq.(\ref{eqFeff}) is generated after an integration over the phonon fields, the analog of $J(x)$
	in our model is the interaction between the $\partial h$ terms, which turn out to be $\mathcal{R}$.
	
	To test if the range of $\mathcal{R}$ is short enough for Hohenberg-Mermin-Wagner's theorem to apply,
	it must be reexpressed in direct space.
	We do not give here the full expression of $\mathcal{R}(x-y)$, but we rather analyze the following elementary
	block:
	\begin{equation}
		P^T_{ab}(q)P_{cd}^T(q)=\delta_{ab}\delta_{cd}-\delta_{ab}\frac{q_cq_d}{q^2}-\delta_{cd}\frac{q_aq_b}
		{q^2}+\frac{q_aq_bq_cq_d}{q^4}\ .
	\end{equation}
	Each term can be expressed in direct space thanks to the following formula of the Fourier transform
	of power laws (see, for example, Ref. \cite{Kleinert01}):
	\begin{equation}
		\frac{1}{\big(p^2\big)^a}=\frac{1}{4^a\pi^\frac{D}{2}}\frac{\Gamma\Big({\small\frac{D}{2}-a}\Big)}
		{\Gamma(a)}\int_x\frac{e^{ip.x}}{\big(x^2\big)^{\frac{D}{2}-a}}\ ,
	\end{equation}
	which finally gives:
	\begin{equation}
	\label{eq123}
		\begin{split}
			& \int_q\delta_{ab}\delta_{cd}e^{i\,q.(x-y)}=\delta_{ab}\delta_{cd}
			\,\delta^{(D)}(x-y)\\
			& \int_q\frac{q_aq_b}{q^2}e^{i\,q.(x-y)}=\frac{\delta_{ab}}{2\pi|x-y|^2}
			-\frac{(x_a-y_a)(x_b-y_b)}{\pi|x-y|^4}\\
			& \int_q\frac{q_aq_bq_cq_d}{q^4}e^{i\,q.(x-y)}=\frac{X_{abcd}}{4\pi|x-y|
			^2}-\frac{Y_{abcd}(x-y)}{2\pi|x-y|^4}\\
			&\quad+\frac{2(x_a-y_a)(x_b-y_b)(x_c-y_c)(x_d-y_d)}{\pi|x-y|^6}\ ,
		\end{split}
	\end{equation}
	with the following tensors being defined:
	\begin{equation}
	\label{eqXY}
		\begin{split}
			& 	X_{\mu\nu\rho\sigma}=\delta_{\mu\nu}\delta_{\rho\sigma}+\delta_{\mu\rho}\delta_{\nu\sigma}
			+\delta_{\mu\sigma}\delta_{\nu\rho}\ ,\\
			& Y_{\mu\nu\rho\sigma}(\vec{x})=x_\mu x_\nu\delta_{\rho\sigma}+x_\mu x_\rho\delta_{\nu\sigma}
			+x_\mu x_\sigma\delta_{\nu\rho}\\
			&\qquad\qquad+x_\nu x_\rho\delta_{\mu\sigma}+x_\nu x_\sigma\delta_{\mu\rho}
			+x_\rho x_\sigma\delta_{\mu\nu}\ .
		\end{split}
	\end{equation}
	Among the different terms of Eq.(\ref{eq123}), only the first one is local.
	The other ones are not integrable over the membrane's internal space once multiplied by $(x-y)^2$.
	Hence, Hohenberg-Mermin-Wagner's theorem does not apply, and the orientational order in crystalline
	membranes can be long-range.
	
	In the previous argument, it is the non-local structure of the effective interaction vertex between
	flexurons that is at the origin of the stabilization of long range order in two dimensions.
	In light of our analysis of the spontaneous symmetry-breaking pattern Eq. (\ref{eqM3}), we
	can add the following: in the flat phase, even if flexurons are the modes that dominate in the
	infrared limit, they are not the only important fluctuation modes.
	In particular, acoustic phonons carry a non-local effective interaction between flexurons at various locations
	in the flat phase's plane.
	The mechanism in Eq. (\ref{eqM3}) moreover guarantees that the phonons are Goldstone modes, i.e.,
	thanks to the massless nature they posses by symmetry, they are not efficiently screened out
	at large distances.
	Thus, the effective interaction they carry is a true long range one, which explains why
	Hohenberg-Mermin-Wagner's theorem does not apply, and the flat phase is robust against thermal
	fluctuations.

\section{The origin of long-range order in the flat phase}

	In the previous sections, we identified two main differences between the third mechanism Eq. (\ref{eqM3})
	and the two other ones Eq. (\ref{eqM1}) and Eq. (\ref{eqM2}).
	In order to refine our comprehension of the necessary ingredients to generate a stable order
	phase in the system, we propose to analyze a fourth mechanism.
	
	Whenever a crystalline membrane undergoes a strong enough stress, it buckles.
	 This buckling phenomenon can be understood as a (second-order) phase transition 
	 \cite{Aronovitz89,Guitter89,Roldan11,Kosmrlj16,Gornyi17} : depending on the type of applied forces,
	 the membrane will either become buckled if compressed -- in this state, the membrane appears as
	 a mosaic of locally flat domains the orientation of which is random \cite{Guitter89} -- or
	 overstretched if sufficiently dilated.
	 Because our main concern is the origin of long-range order in two-dimensional systems, we will
	 focus here on the overstretched phase in which a strong long-range orientational order is present.
	 
	In the overstretched phase, the external stress screens the effects of flexurons \cite{Roldan11,Burmistrov18}
	which become energetically disfavored.
	As a result, the massless infrared spectrum of overstretched membranes contains only type-A
	Goldstone bosons.
	
	Macroscopically, the shape of the overstretched membrane is still flat, with weaker height fluctuations
	compared to the flat phase.
	Microscopically, the stress dominates the local rearrangement of atoms, and the ancient lattice positional order is broken.
	Contrary to the flat phase examined in Sec. \ref{secII}, the ground state in the overstretched phase cannot be uniquely
	characterized by its metric \cite{David88}.
	Indeed, the local arrangement of atoms depends both on the intrinsic properties of the material (captured by $\zeta$) and the external stress.
	Consequently, the previous argument that allowed us to disentangle mixed rotations and external translations
	does not hold anymore: the situation is analogous to that of Fig. \ref{figLM} rather than Fig. \ref{figRotMix}.
	To avoid confusion, and to ensure the disentanglement between the action of translations and mixed rotations on the ground state,
	the symmetry-breaking mechanism will be written as :
	\begin{equation}
	\label{eqM4}
		\text{\underline{\textbf{Mechanism 4:}}} \quad \text{ISO}(d)\rightarrow\text{SO}(d-D)\ .
	\end{equation}
	The set broken generators is the same as for the third mechanism, but now the unit of length at the surface
	of the membrane is determined by the applied stress rather than the sole extension parameter $\zeta$.
	The set of independent broken generators hence reduces to the $d$ translations, so that
	\begin{equation}
		\text{dim}(G/H)=d\,,\quad\rho=0\ ,
	\end{equation}
	and the counting rule Eq. (\ref{eqWata}) leads to:
	\begin{equation}
	\label{eqG4}
		\left\{\begin{split}
			&n_A=d\\
			&n_B=0\ ,
		\end{split}\right.
	\end{equation}
	namely $D$ type-A acoustic phonons and $d-D$ type-A flexurons, as expected.
	
	Finally, mechanism 4 Eq. (\ref{eqM4}) provides an example in which long-range orientational order can be 
	maintained in two dimensions without requiring the help of type-B Goldstone bosons.
	Indeed in the overstretched phase, the previous argument with regard to the Hohenberg-Mermin-Wagner's
	theorem still holds: despite the fact that the flexurons are screened by the stress, they remain
	Goldstone bosons, and an effective theory of interacting flexurons can still be built, in which
	the phonons, which are also massless, carry an effective long-range interaction between flexurons.

	Note, however, that in the ground state of overstretched membranes the local pseudo-ordering persists, which causes the	
	breaking of the group isometries inside
	the membrane's plane, which is in strong contrast with the second mechanism Eq. (\ref{eqM2}), 
	in which the microcopic constituents are free to move and no phonon is generated (and therefore
	no long range interaction can occur and the ordered phase is destroyed by the thermal fluctuations).
	
	A remaining question involves the role of the type-B Goldstone bosons.
	To address it, we must compare the mechanisms 3 Eq. (\ref{eqM3}) and 4 Eq. (\ref{eqM4}), which
	differ only by the dispersion relation of the flexurons.
	The most striking difference between the flat phase and the overstretched phase is the presence
	in the former of a strong anomalous exponent $\eta\simeq0.85$ \cite{Nelson04,Kownacki09,Essafi14,Los16} , at the origin of
	the highly anharmonic behavior of the thermal fluctuations, which leads to many unusual effects,
	as well as a modified elasticity theory, which is in total contrast with the overstretched
	phase in which conventional elasticity is restored and $\eta=0$ 
	(see \cite{Roldan11,Kosmrlj16,Burmistrov18} and \cite{Coquand19} for a comparative study).
	The role of the type-B Goldstone bosons is thus probably related to the generation of such an
	anharmonic behavior and unusual scaling relations.

	To sum up on the Goldstone physics, we have analyzed various non-trivial symmetry breaking patterns related to the physics of
	crystalline membranes, which have highlighted a number of general features of the physics of Goldstone
	modes in theories without Lorentz invariance.
	First, mechanism 1 Eq. (\ref{eqM1}) illustrates the fact that whenever different broken symmetry generators
	generate linearly dependent transformations of the ground state, the total number of associated
	Goldstone modes is smaller than the total number of broken generators, which is the main lesson
	of reference \cite{Low02}.
	This feature is quite common in theories presenting spacetime symmetry breakings, since the action
	of rotations and translations are rarely independent (it is much more difficult to see this if an internal
	symmetry group is broken).
	Mechanism 2 Eq. (\ref{eqM2}) teaches us the importance of the underlying microscopics, even in continuum
	theories.
	Indeed, taking into account the mere overall shape of the membrane leads to a spontaneous symmetry
	breaking mechanism without phonons, which is much more sensitive to thermal fluctuations (as such
	a system cannot sustain long range order in two dimensions).
	The comparison between mechanisms 3 Eq. (\ref{eqM3}) and 4 Eq. (\ref{eqM4}) shows that the presence
	of two different types of interacting Goldstone modes is a sufficient condition to
	generate a stable ordered phase in two dimensions, as whenever one can reexpress the theory as
	an effective theory of a single mode, the effective interaction carried by the second Goldstone mode
	must be long range, because of its massless character.
	Having two different types of Goldstone modes, however, requires particular patterns of symmetry breaking.
	Finally, we also showed the necessity of having a non-trivial algebra of independently acting
	broken generators to generate type-B Goldstone bosons, which can be seen as an obvious consequence of the
	Goldstone counting rule
	Eq. (\ref{eqWata}), but which we have shown is not so easy to achieve in practice.
	We also give hints at the possible link between the presence of such type-B Goldstone bosons and unusual
	scaling behaviors in the ordered phase, related to the generation of a non-trivial field anomalous
	dimension $\eta$ .

\section{Conclusion}
	
	To conclude, the corrections of the symmetry breaking mechanism at the origin of the flat phase teach us two main lessons on the physics of crystalline
	membranes.
	First, acoustic phonons cannot be overlooked, even though crytalline order is broken by thermal fluctuations, and by the fact that they are subdominant at large
	distances.
	It is all the more important that the presence of a massless effective interaction carrier is required to ensure that long-range orientational order is not
	destroyed by fluctuations.
	This is why, genuine bidimensional crystals or fluid membranes, in which only phonons alone, or flexurons alone survive at large distances, do not present
	any stable ordered phase in two dimensions, but crystalline membranes, which possess both modes, also present long-range order.

	Second, the origin of the flat phase anomalous scaling laws can be traced back to a delicate geometrical interplay between the intrinsic properties of the material,
	and its embedding in the three dimensional space.
	In the presence of a strong enough external stress field -- which drives the system into the overstretched phase -- this subtle balance is broken,
	and conventional elasticity is restored, probably due to the absence of type-B Goldstone bosons.

	The key to understanding all these results is the Goldstone counting rule Eq. (\ref{eqWata}).
	We hope that our work will motivate further studies in the context of condensed matter physics, in which Lorentz invariance
	is frequently absent, spacetime symmetries are often at play, and therefore such tools are certainly of interest.

\section*{Acknowledgements}

	The author thanks D. Mouhanna for useful discussions and a careful reading of this manuscript.

\bibliography{Goldpaper.bib}

\begin{thebibliography}{45}
\expandafter\ifx\csname natexlab\endcsname\relax\def\natexlab#1{#1}\fi
\expandafter\ifx\csname bibnamefont\endcsname\relax
  \def\bibnamefont#1{#1}\fi
\expandafter\ifx\csname bibfnamefont\endcsname\relax
  \def\bibfnamefont#1{#1}\fi
\expandafter\ifx\csname citenamefont\endcsname\relax
  \def\citenamefont#1{#1}\fi
\expandafter\ifx\csname url\endcsname\relax
  \def\url#1{\texttt{#1}}\fi
\expandafter\ifx\csname urlprefix\endcsname\relax\def\urlprefix{URL }\fi
\providecommand{\bibinfo}[2]{#2}
\providecommand{\eprint}[2][]{\url{#2}}

\bibitem[{\citenamefont{Peierls}(1934)}]{Peierls34}
\bibinfo{author}{\bibfnamefont{R.}~\bibnamefont{Peierls}},
  \bibinfo{journal}{Helv. Phys. Acta} \textbf{\bibinfo{volume}{7}},
  \bibinfo{pages}{81} (\bibinfo{year}{1934}).

\bibitem[{\citenamefont{Peierls}(1935)}]{Peierls35}
\bibinfo{author}{\bibfnamefont{R.}~\bibnamefont{Peierls}},
  \bibinfo{journal}{Annales de l'Institut Henri Poincar{\'e}}
  \textbf{\bibinfo{volume}{5}}, \bibinfo{pages}{177} (\bibinfo{year}{1935}).

\bibitem[{\citenamefont{Landau}(1937)}]{Landau37}
\bibinfo{author}{\bibfnamefont{L.}~\bibnamefont{Landau}},
  \bibinfo{journal}{JETP (Sov. Phys.)} \textbf{\bibinfo{volume}{7}},
  \bibinfo{pages}{627} (\bibinfo{year}{1937}).

\bibitem[{\citenamefont{Mermin}(1968)}]{Mermin68}
\bibinfo{author}{\bibfnamefont{N.}~\bibnamefont{Mermin}},
  \bibinfo{journal}{Phys. Rev.} \textbf{\bibinfo{volume}{176}},
  \bibinfo{pages}{250} (\bibinfo{year}{1968}).

\bibitem[{\citenamefont{Novoselov et~al.}(2004)\citenamefont{Novoselov, Geim,
  Morozov, Jiang, Zhang, Dubonos, Grigorieva, and Firsov}}]{Novoselov04}
\bibinfo{author}{\bibfnamefont{K.}~\bibnamefont{Novoselov}},
  \bibinfo{author}{\bibfnamefont{A.}~\bibnamefont{Geim}},
  \bibinfo{author}{\bibfnamefont{S.}~\bibnamefont{Morozov}},
  \bibinfo{author}{\bibfnamefont{D.}~\bibnamefont{Jiang}},
  \bibinfo{author}{\bibfnamefont{Y.}~\bibnamefont{Zhang}},
  \bibinfo{author}{\bibfnamefont{S.}~\bibnamefont{Dubonos}},
  \bibinfo{author}{\bibfnamefont{I.}~\bibnamefont{Grigorieva}},
  \bibnamefont{and} \bibinfo{author}{\bibfnamefont{A.}~\bibnamefont{Firsov}},
  \bibinfo{journal}{Science} \textbf{\bibinfo{volume}{306}},
  \bibinfo{pages}{666} (\bibinfo{year}{2004}).

\bibitem[{\citenamefont{Novoselov}(2011)}]{Novoselov10}
\bibinfo{author}{\bibfnamefont{K.}~\bibnamefont{Novoselov}},
  \bibinfo{journal}{International Journal of Modern Physics B}
  \textbf{\bibinfo{volume}{25}}, \bibinfo{pages}{4081} (\bibinfo{year}{2011}).

\bibitem[{\citenamefont{Rold{\'a}n et~al.}(2017)\citenamefont{Rold{\'a}n,
  Chirolli, Prada, Silva-Guill{\'e}n, San-Jose, and Guinea}}]{Roldan17}
\bibinfo{author}{\bibfnamefont{R.}~\bibnamefont{Rold{\'a}n}},
  \bibinfo{author}{\bibfnamefont{L.}~\bibnamefont{Chirolli}},
  \bibinfo{author}{\bibfnamefont{E.}~\bibnamefont{Prada}},
  \bibinfo{author}{\bibfnamefont{J.}~\bibnamefont{Silva-Guill{\'e}n}},
  \bibinfo{author}{\bibfnamefont{P.}~\bibnamefont{San-Jose}}, \bibnamefont{and}
  \bibinfo{author}{\bibfnamefont{F.}~\bibnamefont{Guinea}},
  \bibinfo{journal}{Chem. Soc. Rev.} \textbf{\bibinfo{volume}{46}},
  \bibinfo{pages}{4387} (\bibinfo{year}{2017}).

\bibitem[{\citenamefont{Aronovitz and Lubensky}(1988)}]{Aronovitz88}
\bibinfo{author}{\bibfnamefont{J.}~\bibnamefont{Aronovitz}} \bibnamefont{and}
  \bibinfo{author}{\bibfnamefont{T.}~\bibnamefont{Lubensky}},
  \bibinfo{journal}{Phys. Rev. Lett.} \textbf{\bibinfo{volume}{60}},
  \bibinfo{pages}{2634} (\bibinfo{year}{1988}).

\bibitem[{\citenamefont{Aronovitz et~al.}(1989)\citenamefont{Aronovitz,
  Golubovic, and Lubensky}}]{Aronovitz89}
\bibinfo{author}{\bibfnamefont{J.}~\bibnamefont{Aronovitz}},
  \bibinfo{author}{\bibfnamefont{L.}~\bibnamefont{Golubovic}},
  \bibnamefont{and} \bibinfo{author}{\bibfnamefont{T.}~\bibnamefont{Lubensky}},
  \bibinfo{journal}{J. Phys. (Paris)} \textbf{\bibinfo{volume}{50}},
  \bibinfo{pages}{609} (\bibinfo{year}{1989}).

\bibitem[{\citenamefont{Meyer et~al.}(2007)\citenamefont{Meyer, Geim,
  Katsnelson, Novoselov, Booth, and Roth}}]{Meyer07}
\bibinfo{author}{\bibfnamefont{J.}~\bibnamefont{Meyer}},
  \bibinfo{author}{\bibfnamefont{A.}~\bibnamefont{Geim}},
  \bibinfo{author}{\bibfnamefont{M.}~\bibnamefont{Katsnelson}},
  \bibinfo{author}{\bibfnamefont{K.}~\bibnamefont{Novoselov}},
  \bibinfo{author}{\bibfnamefont{T.}~\bibnamefont{Booth}}, \bibnamefont{and}
  \bibinfo{author}{\bibfnamefont{S.}~\bibnamefont{Roth}},
  \bibinfo{journal}{nature letters} \textbf{\bibinfo{volume}{446}},
  \bibinfo{pages}{60} (\bibinfo{year}{2007}).

\bibitem[{\citenamefont{Nelson et~al.}(2004)\citenamefont{Nelson, Piran, and
  Weinberg}}]{Nelson04}
\bibinfo{author}{\bibfnamefont{D.}~\bibnamefont{Nelson}},
  \bibinfo{author}{\bibfnamefont{T.}~\bibnamefont{Piran}}, \bibnamefont{and}
  \bibinfo{author}{\bibfnamefont{S.}~\bibnamefont{Weinberg}},
  \emph{\bibinfo{title}{Proceedings of the Fifth Jerusalem Winter School for
  Theoretical Physics}} (\bibinfo{publisher}{Word Scientific Singapore},
  \bibinfo{year}{2004}).

\bibitem[{\citenamefont{Guitter et~al.}(1989)\citenamefont{Guitter, David,
  Leibler, and Peliti}}]{Guitter89}
\bibinfo{author}{\bibfnamefont{E.}~\bibnamefont{Guitter}},
  \bibinfo{author}{\bibfnamefont{F.}~\bibnamefont{David}},
  \bibinfo{author}{\bibfnamefont{S.}~\bibnamefont{Leibler}}, \bibnamefont{and}
  \bibinfo{author}{\bibfnamefont{L.}~\bibnamefont{Peliti}},
  \bibinfo{journal}{J. Phys. France} \textbf{\bibinfo{volume}{50}},
  \bibinfo{pages}{1787} (\bibinfo{year}{1989}).

\bibitem[{\citenamefont{Gornyi et~al.}(2017)\citenamefont{Gornyi, Kachorovskii,
  and Mirlin}}]{Gornyi17}
\bibinfo{author}{\bibfnamefont{I.}~\bibnamefont{Gornyi}},
  \bibinfo{author}{\bibfnamefont{V.}~\bibnamefont{Kachorovskii}},
  \bibnamefont{and} \bibinfo{author}{\bibfnamefont{A.}~\bibnamefont{Mirlin}},
  \bibinfo{journal}{2D Material} \textbf{\bibinfo{volume}{4}},
  \bibinfo{pages}{011003} (\bibinfo{year}{2017}).

\bibitem[{\citenamefont{Mermin and Wagner}(1966)}]{Mermin66}
\bibinfo{author}{\bibfnamefont{N.}~\bibnamefont{Mermin}} \bibnamefont{and}
  \bibinfo{author}{\bibfnamefont{H.}~\bibnamefont{Wagner}},
  \bibinfo{journal}{Phys. Rev. Lett.} \textbf{\bibinfo{volume}{17}}
  (\bibinfo{year}{1966}).

\bibitem[{\citenamefont{Hohenberg}(1967)}]{Hohenberg67}
\bibinfo{author}{\bibfnamefont{P.}~\bibnamefont{Hohenberg}},
  \bibinfo{journal}{Phys. Rev.} \textbf{\bibinfo{volume}{158}},
  \bibinfo{pages}{383} (\bibinfo{year}{1967}).

\bibitem[{\citenamefont{Nelson and Peliti}(1987)}]{Nelson87}
\bibinfo{author}{\bibfnamefont{D.}~\bibnamefont{Nelson}} \bibnamefont{and}
  \bibinfo{author}{\bibfnamefont{L.}~\bibnamefont{Peliti}},
  \bibinfo{journal}{J. Phys.} \textbf{\bibinfo{volume}{48}},
  \bibinfo{pages}{1085} (\bibinfo{year}{1987}).

\bibitem[{\citenamefont{Goldstone et~al.}(1962)\citenamefont{Goldstone, Salam,
  and Weinberg}}]{Goldstone62}
\bibinfo{author}{\bibfnamefont{J.}~\bibnamefont{Goldstone}},
  \bibinfo{author}{\bibfnamefont{A.}~\bibnamefont{Salam}}, \bibnamefont{and}
  \bibinfo{author}{\bibfnamefont{S.}~\bibnamefont{Weinberg}},
  \bibinfo{journal}{Phys. Rev.} \textbf{\bibinfo{volume}{27}},
  \bibinfo{pages}{965} (\bibinfo{year}{1962}).

\bibitem[{\citenamefont{Nielsen and Chadha}(1976)}]{Nielsen76}
\bibinfo{author}{\bibfnamefont{H.}~\bibnamefont{Nielsen}} \bibnamefont{and}
  \bibinfo{author}{\bibfnamefont{S.}~\bibnamefont{Chadha}},
  \bibinfo{journal}{Nucl. Phys. B} \textbf{\bibinfo{volume}{105}},
  \bibinfo{pages}{445} (\bibinfo{year}{1976}).

\bibitem[{\citenamefont{Brauner}(2010)}]{Brauner10}
\bibinfo{author}{\bibfnamefont{T.}~\bibnamefont{Brauner}},
  \bibinfo{journal}{Symmetry} \textbf{\bibinfo{volume}{2}},
  \bibinfo{pages}{609} (\bibinfo{year}{2010}).

\bibitem[{\citenamefont{Watanabe and Brauner}(2011)}]{Watanabe11}
\bibinfo{author}{\bibfnamefont{H.}~\bibnamefont{Watanabe}} \bibnamefont{and}
  \bibinfo{author}{\bibfnamefont{T.}~\bibnamefont{Brauner}},
  \bibinfo{journal}{Phys. Rev. D} \textbf{\bibinfo{volume}{84}},
  \bibinfo{pages}{125013} (\bibinfo{year}{2011}).

\bibitem[{\citenamefont{Watanabe and Murayama}(2012)}]{Watanabe12}
\bibinfo{author}{\bibfnamefont{H.}~\bibnamefont{Watanabe}} \bibnamefont{and}
  \bibinfo{author}{\bibfnamefont{H.}~\bibnamefont{Murayama}},
  \bibinfo{journal}{Phys. Rev. Lett.} \textbf{\bibinfo{volume}{108}},
  \bibinfo{pages}{251602} (\bibinfo{year}{2012}).

\bibitem[{\citenamefont{Watanabe and Murayama}(2014)}]{Watanabe14}
\bibinfo{author}{\bibfnamefont{H.}~\bibnamefont{Watanabe}} \bibnamefont{and}
  \bibinfo{author}{\bibfnamefont{H.}~\bibnamefont{Murayama}},
  \bibinfo{journal}{Phys. Rev. X} \textbf{\bibinfo{volume}{4}},
  \bibinfo{pages}{031057} (\bibinfo{year}{2014}).

\bibitem[{\citenamefont{Hidaka}(2013)}]{Hidaka13}
\bibinfo{author}{\bibfnamefont{Y.}~\bibnamefont{Hidaka}},
  \bibinfo{journal}{Phys. Rev. Lett.} \textbf{\bibinfo{volume}{110}},
  \bibinfo{pages}{091601} (\bibinfo{year}{2013}).

\bibitem[{\citenamefont{Zanusso}(2014)}]{Zanusso14}
\bibinfo{author}{\bibfnamefont{O.}~\bibnamefont{Zanusso}},
  \bibinfo{journal}{Phys. Rev. E} \textbf{\bibinfo{volume}{90}},
  \bibinfo{pages}{052110} (\bibinfo{year}{2014}).

\bibitem[{\citenamefont{de~Gennes and Taupin}(1982)}]{DeGennes82}
\bibinfo{author}{\bibfnamefont{P.}~\bibnamefont{de~Gennes}} \bibnamefont{and}
  \bibinfo{author}{\bibfnamefont{C.}~\bibnamefont{Taupin}},
  \bibinfo{journal}{J. Phys. Chem.} \textbf{\bibinfo{volume}{86}},
  \bibinfo{pages}{2294} (\bibinfo{year}{1982}).

\bibitem[{\citenamefont{Nicolis and Piazza}(2013)}]{Nicolis13}
\bibinfo{author}{\bibfnamefont{A.}~\bibnamefont{Nicolis}} \bibnamefont{and}
  \bibinfo{author}{\bibfnamefont{F.}~\bibnamefont{Piazza}},
  \bibinfo{journal}{Phys. Rev. Lett.} \textbf{\bibinfo{volume}{110}},
  \bibinfo{pages}{011602} (\bibinfo{year}{2013}).

\bibitem[{\citenamefont{Nicolis et~al.}(2013)\citenamefont{Nicolis, Penco,
  Piazza, and Rosen}}]{Nicolis13a}
\bibinfo{author}{\bibfnamefont{A.}~\bibnamefont{Nicolis}},
  \bibinfo{author}{\bibfnamefont{R.}~\bibnamefont{Penco}},
  \bibinfo{author}{\bibfnamefont{F.}~\bibnamefont{Piazza}}, \bibnamefont{and}
  \bibinfo{author}{\bibfnamefont{R.}~\bibnamefont{Rosen}},
  \bibinfo{journal}{Journal of High Energy Physics}
  \textbf{\bibinfo{volume}{2013}}, \bibinfo{pages}{55} (\bibinfo{year}{2013}).

\bibitem[{\citenamefont{Low and Manohar}(2002)}]{Low02}
\bibinfo{author}{\bibfnamefont{I.}~\bibnamefont{Low}} \bibnamefont{and}
  \bibinfo{author}{\bibfnamefont{A.}~\bibnamefont{Manohar}},
  \bibinfo{journal}{Phys. Rev. Lett.} \textbf{\bibinfo{volume}{88}},
  \bibinfo{pages}{101602} (\bibinfo{year}{2002}).

\bibitem[{\citenamefont{Kharuk and Shkerin}(2018)}]{Kharuk18}
\bibinfo{author}{\bibfnamefont{I.}~\bibnamefont{Kharuk}} \bibnamefont{and}
  \bibinfo{author}{\bibfnamefont{A.}~\bibnamefont{Shkerin}},
  \bibinfo{journal}{Phys. Rev. D} \textbf{\bibinfo{volume}{98}},
  \bibinfo{pages}{125016} (\bibinfo{year}{2018}).

\bibitem[{\citenamefont{Landau et~al.}(1990)\citenamefont{Landau, Lifschitz,
  and Kosevich}}]{Landau90}
\bibinfo{author}{\bibfnamefont{L.}~\bibnamefont{Landau}},
  \bibinfo{author}{\bibfnamefont{E.}~\bibnamefont{Lifschitz}},
  \bibnamefont{and} \bibinfo{author}{\bibfnamefont{A.}~\bibnamefont{Kosevich}},
  \emph{\bibinfo{title}{Physique th{\'e}orique}}, vol. \bibinfo{volume}{7,
  Th{\'e}orie de l'{\'e}lasticit{\'e}} (\bibinfo{publisher}{{\'E}ditions Mir},
  \bibinfo{year}{1990}).

\bibitem[{\citenamefont{Watanabe and Murayama}(2013)}]{Watanabe13}
\bibinfo{author}{\bibfnamefont{H.}~\bibnamefont{Watanabe}} \bibnamefont{and}
  \bibinfo{author}{\bibfnamefont{H.}~\bibnamefont{Murayama}},
  \bibinfo{journal}{Phys. Rev. Lett.} \textbf{\bibinfo{volume}{110}},
  \bibinfo{pages}{181601} (\bibinfo{year}{2013}).

\bibitem[{\citenamefont{Beekman et~al.}(2017)\citenamefont{Beekman, Nissinen,
  Wu, and Zaanen}}]{Beekman17a}
\bibinfo{author}{\bibfnamefont{A.~J.} \bibnamefont{Beekman}},
  \bibinfo{author}{\bibfnamefont{J.}~\bibnamefont{Nissinen}},
  \bibinfo{author}{\bibfnamefont{K.}~\bibnamefont{Wu}}, \bibnamefont{and}
  \bibinfo{author}{\bibfnamefont{J.}~\bibnamefont{Zaanen}},
  \bibinfo{journal}{Phys. Rev. B} \textbf{\bibinfo{volume}{96}},
  \bibinfo{pages}{165115} (\bibinfo{year}{2017}).

\bibitem[{\citenamefont{Schwarz}(1971)}]{Schwarz71}
\bibinfo{author}{\bibfnamefont{F.}~\bibnamefont{Schwarz}},
  \bibinfo{journal}{Annales de l'Institut Henri Poincar{\'e} Physique
  th{\'e}orique} \textbf{\bibinfo{volume}{15}}, \bibinfo{pages}{15}
  (\bibinfo{year}{1971}).

\bibitem[{\citenamefont{Canham}(1970)}]{Canham70}
\bibinfo{author}{\bibfnamefont{P.}~\bibnamefont{Canham}}, \bibinfo{journal}{J.
  Theoret. Biol.} \textbf{\bibinfo{volume}{26}}, \bibinfo{pages}{61}
  (\bibinfo{year}{1970}).

\bibitem[{\citenamefont{Helfrich}(1973)}]{Helfrich73}
\bibinfo{author}{\bibfnamefont{W.}~\bibnamefont{Helfrich}},
  \bibinfo{journal}{Z. Naturforsch.} \textbf{\bibinfo{volume}{28 c}},
  \bibinfo{pages}{693} (\bibinfo{year}{1973}).

\bibitem[{\citenamefont{David and Guitter}(1988)}]{David88}
\bibinfo{author}{\bibfnamefont{F.}~\bibnamefont{David}} \bibnamefont{and}
  \bibinfo{author}{\bibfnamefont{E.}~\bibnamefont{Guitter}},
  \bibinfo{journal}{Europhys. Lett.} \textbf{\bibinfo{volume}{5}},
  \bibinfo{pages}{709} (\bibinfo{year}{1988}).

\bibitem[{\citenamefont{LeDoussal and Radzihovsky}(1992)}]{LeDoussal92}
\bibinfo{author}{\bibfnamefont{P.}~\bibnamefont{LeDoussal}} \bibnamefont{and}
  \bibinfo{author}{\bibfnamefont{L.}~\bibnamefont{Radzihovsky}},
  \bibinfo{journal}{Phys. Rev. Lett.} \textbf{\bibinfo{volume}{69}},
  \bibinfo{pages}{1209} (\bibinfo{year}{1992}).

\bibitem[{\citenamefont{Kleinert and Schulte-Frohlinde}(2001)}]{Kleinert01}
\bibinfo{author}{\bibfnamefont{H.}~\bibnamefont{Kleinert}} \bibnamefont{and}
  \bibinfo{author}{\bibfnamefont{V.}~\bibnamefont{Schulte-Frohlinde}},
  \emph{\bibinfo{title}{Critical Properties of $\phi^4$-theories}}
  (\bibinfo{publisher}{World Scientific, Singapore}, \bibinfo{year}{2001}).

\bibitem[{\citenamefont{Rold{\`a}n et~al.}(2011)\citenamefont{Rold{\`a}n,
  Fasolino, Zakharchenko, and Katsnelson}}]{Roldan11}
\bibinfo{author}{\bibfnamefont{R.}~\bibnamefont{Rold{\`a}n}},
  \bibinfo{author}{\bibfnamefont{A.}~\bibnamefont{Fasolino}},
  \bibinfo{author}{\bibfnamefont{K.}~\bibnamefont{Zakharchenko}},
  \bibnamefont{and}
  \bibinfo{author}{\bibfnamefont{M.}~\bibnamefont{Katsnelson}},
  \bibinfo{journal}{Phys. Rev B} \textbf{\bibinfo{volume}{83}},
  \bibinfo{pages}{174104} (\bibinfo{year}{2011}).

\bibitem[{\citenamefont{Ko{$\check{\text{s}}$}mrlj and
  Nelson}(2016)}]{Kosmrlj16}
\bibinfo{author}{\bibfnamefont{A.}~\bibnamefont{Ko{$\check{\text{s}}$}mrlj}}
  \bibnamefont{and} \bibinfo{author}{\bibfnamefont{D.}~\bibnamefont{Nelson}},
  \bibinfo{journal}{Phys. Rev. B} \textbf{\bibinfo{volume}{93}},
  \bibinfo{pages}{125431} (\bibinfo{year}{2016}).

\bibitem[{\citenamefont{Burmistrov et~al.}(2018)\citenamefont{Burmistrov,
  Gornyi, Kachorovskii, Katsnelson, Los, and Mirlin}}]{Burmistrov18}
\bibinfo{author}{\bibfnamefont{I.}~\bibnamefont{Burmistrov}},
  \bibinfo{author}{\bibfnamefont{I.}~\bibnamefont{Gornyi}},
  \bibinfo{author}{\bibfnamefont{V.}~\bibnamefont{Kachorovskii}},
  \bibinfo{author}{\bibfnamefont{M.}~\bibnamefont{Katsnelson}},
  \bibinfo{author}{\bibfnamefont{J.}~\bibnamefont{Los}}, \bibnamefont{and}
  \bibinfo{author}{\bibfnamefont{A.}~\bibnamefont{Mirlin}},
  \bibinfo{journal}{Phys. Rev. B} \textbf{\bibinfo{volume}{97}},
  \bibinfo{pages}{125402} (\bibinfo{year}{2018}).

\bibitem[{\citenamefont{Kownacki and Mouhanna}(2009)}]{Kownacki09}
\bibinfo{author}{\bibfnamefont{J.}~\bibnamefont{Kownacki}} \bibnamefont{and}
  \bibinfo{author}{\bibfnamefont{D.}~\bibnamefont{Mouhanna}},
  \bibinfo{journal}{Phys. Rev. E} \textbf{\bibinfo{volume}{79}},
  \bibinfo{pages}{040101} (\bibinfo{year}{2009}).

\bibitem[{\citenamefont{Essafi et~al.}(2014)\citenamefont{Essafi, Kownacki, and
  Mouhanna}}]{Essafi14}
\bibinfo{author}{\bibfnamefont{K.}~\bibnamefont{Essafi}},
  \bibinfo{author}{\bibfnamefont{J.}~\bibnamefont{Kownacki}}, \bibnamefont{and}
  \bibinfo{author}{\bibfnamefont{D.}~\bibnamefont{Mouhanna}},
  \bibinfo{journal}{Phys. Rev. E} \textbf{\bibinfo{volume}{89}},
  \bibinfo{pages}{042101} (\bibinfo{year}{2014}).

\bibitem[{\citenamefont{Los et~al.}(2016)\citenamefont{Los, Fasolino, and
  Katsnelson}}]{Los16}
\bibinfo{author}{\bibfnamefont{J.}~\bibnamefont{Los}},
  \bibinfo{author}{\bibfnamefont{A.}~\bibnamefont{Fasolino}}, \bibnamefont{and}
  \bibinfo{author}{\bibfnamefont{M.}~\bibnamefont{Katsnelson}},
  \bibinfo{journal}{Phys. Rev. Lett.} \textbf{\bibinfo{volume}{116}},
  \bibinfo{pages}{015901} (\bibinfo{year}{2016}).

\bibitem[{\citenamefont{Coquand and Mouhanna}(2019)}]{Coquand19}
\bibinfo{author}{\bibfnamefont{O.}~\bibnamefont{Coquand}} \bibnamefont{and}
  \bibinfo{author}{\bibfnamefont{D.}~\bibnamefont{Mouhanna}},
  \bibinfo{journal}{to be published}  (\bibinfo{year}{2019}).

\end{thebibliography}

\end{document}